\documentclass[12pt]{article}
\usepackage{amsmath,amssymb,amscd}
\newcommand{\x}{\times}
\renewcommand{\>}{\rangle}
\renewcommand{\i}{\infty}
\newcommand{\p}{\partial}

\renewcommand{\d}{\delta}
\newcommand{\D}{\Delta}
\newcommand{\s}{\sigma}
\renewcommand{\th}{\theta}
\renewcommand{\o}{\omega}

\renewcommand{\frac}{\tfrac}

\newcommand{\perm}{\mathop{\mathrm{perm}}\nolimits}
\newcommand{\open}{\mathop{\mathrm{open}}\nolimits}
\renewcommand{\mod}{\mathop{\mathrm{mod}}\nolimits}

\newcommand{\lcol}{:\!\!}
\newcommand{\rcol}{\!\!:}
\newcommand{\hA}{{\hat A}}

\newcommand{\hB}{{\hat B}}
\newcommand{\hG}{{\hat G}}
\newcommand{\hH}{{\hat H}}
\newcommand{\hJ}{{\hat J}}
\newcommand{\hL}{{\hat L}}
\newcommand{\hQ}{{\hat Q}}
\newcommand{\hS}{{\hat S}}
\newcommand{\hX}{{\hat X}}
\newcommand{\ha}{{\hat a}}
\newcommand{\hb}{{\hat b}}
\newcommand{\hc}{{\hat c}}
\newcommand{\hh}{{\hat h}}
\newcommand{\hj}{{\hat j}}
\newcommand{\hk}{{\hat k}}
\newcommand{\hl}{{\hat l}}
\newcommand{\hm}{{\hat m}}
\newcommand{\bn}{{\bar n}}
\newcommand{\bu}{{\bar u}}
\newcommand{\hbj}{{\hat{\bar j}}}
\newcommand{\bhj}{{\bar{\hat j}}}

\newcommand{\hpsi}{{\hat \psi}}
\newcommand{\bpsi}{{\bar \psi}}
\newcommand{\hbpsi}{{\hat{\bar \psi}}}
\newcommand{\hth}{{\hat \th}}
\newcommand{\hD}{{\hat \D}}

\newcommand{\bZ}{{\mathbb Z}}
\newcommand{\cA}{{\mathcal A}}
\newcommand{\cG}{{\mathcal G}}
\newcommand{\cO}{{\mathcal O}}

\title{The Orbifold-String Theories of Permutation-Type: \\
I.  One Twisted BRST per Cycle per Sector}
\author{M.~B.~Halpern\thanks{halpern@physics.berkeley.edu} \\
Department of Physics \\
University of California \\
Berkeley, CA\ \ 94720, USA}
\date{}

\begin{document}

\maketitle

\begin{abstract}
\noindent We resume our discussion of the new orbifold-string theories of 
permutation-type, focusing in the present series on the algebraic 
formulation of the general bosonic prototype and especially the target space-times 
of the theories. In this first paper of the series, we construct one 
twisted BRST system for each cycle $j$ in each twisted sector $\sigma$ of the 
general case, verifying in particular the previously-conjectured algebra 
$[\hat{Q}_{i}(\sigma),\hat{Q}_{j}(\sigma)]_{+} =0$ of the BRST charges.
The BRST systems then imply a set of extended physical-state conditions for 
the matter of each cycle at cycle central
charge $\hc_j(\s) = 26f_j(\s)$, where $f_j(\s)$ is the length of cycle~$j$.
\end{abstract}

\clearpage
\section*{Table of Contents}

{\bf \noindent
1\quad Introduction \hfill 3

\bigskip
\noindent
2\quad BRST in the Orbifold Program \hfill 6 

\bigskip
\noindent
3\quad Operator-Product Form of the Twisted BRST's \hfill 9 

\bigskip
\noindent
4\quad The Mode Algebras of Twisted Sector $\s$ \hfill 12

\bigskip
\noindent
5\quad One BRST Charge per Cycle per Sector \hfill 16

\bigskip
\noindent
6\quad The Physical States of Cycle $j$ in Sector $\s$ \hfill 17

\bigskip
\noindent
7\quad Mode-Ordered Form of the Twisted Ghost Systems \hfill 18

\bigskip
\noindent
8\quad The Extended Physical-State Conditions \hfill 20

\bigskip
\noindent
9\quad Conclusions \hfill 24

\bigskip
\noindent
Appendix A.  Fermi Statistics for $\hc$'s \hfill 26

\bigskip
\noindent
Appendix B.  Extended Actions and Generalized Orientation Orbifolds \hfill 27}

\clearpage
\section{Introduction}

The orbifold program [1-15] systematically constructs the twisted sectors of the orbifold conformal 
field theory $A(H)/H$ from any current-algebraic conformal field theory $A(H)$ with a finite symmetry group $H$.  
A short review of the program at the conformal-field-theoretic level is included in Ref.~[10].

Opening a new, more phenomenological chapter of the program, I have recently proposed [16-20] that
a very simple subset of these orbifolds, called the {\em orbifold-string 
theories of permutation-type}, can provide {\em new
physical string theories} at multiples of conventional critical central charges.  

This class of theories begins by choosing
the conformal field theory $A(H)$ to be a set of copies of any critical string (hence only Abelian currents), 
and there are many of these, including the {\em bosonic prototypes} [16]
\[
\frac {U(1)^{26K}}{H_+} = \frac {U(1)_1^{26} \x \dots \x U(1)_K^{26}}{H_+},\ \ H_+ \subset H(\perm)_K \x H'_{26} \eqno(1.1a)
\]
\[
\frac {U(1)^{26}}{H_-} = \frac {U(1)_L^{26} \x U(1)_R^{26}}{H_-},\ \ H_- \subset \bZ_2(w.s.) \x H'_{26} \eqno(1.1b)
\]
\[
[ \frac {U(1)^{26K}}{H_+}]_{\open} \eqno(1.1c)
\]
and orbifold-superstring generalizations of these.  In this notation, $U(1)^{26}$ is the critical bosonic closed string at central charge 26
and $U(1)_{L,R}^{26}$ are the left- and right-movers of $U(1)^{26}$.  The automorphism group $H(\perm)_K$ is 
any permutation group on $K$ elements (the copies), the non-trivial element of $\bZ_2(w.s.)$ is the exchange $L \leftrightarrow R$ of
the chiral components, and the automorphism group $H'_{26}$ operates uniformly on each copy of $U(1)^{26}$, including $U(1)_L^{26}$ and $U(1)_R^{26}$.
The divisors $H_{\pm}$ can be any subgroup of the direct products shown.

The sets shown in Eq.~(1.1a) are called the {\em generalized permutation 
orbifolds} [9,15,16,18,20], with every sector $\s$  being a twisted closed string
at central charge $\hc(\s) = 26K$.  The second sets in Eq.~(1.1b) are the {\em orientation orbifolds} [12,13,15-19], with an equal number of
twisted open strings at $\hc(\sigma) = 52$ and twisted closed strings at $\hc(\sigma) = 
26$. The closed-string sectors of the orientation orbifolds form the 
ordinary space-time orbifolds $U(1)^{26}/H'_{26}$, which we will discuss 
separately elsewhere. The last sets in Eq.~(1.1c) are the so-called ``open-string
generalized permutation orbifolds'' with all sectors at $\hc(\s) = 26K$. These sets are constructed along with their
branes in Refs.~[15,16] from the left-movers
of the generalized permutation orbifolds.  The $\hc = 52$ open-string sectors of the orientation orbifolds are included among the $T$-dual 
families of $[U(1)^{52}/H_+]_{\open}$, but this is the only case in the third set whose closure to closed strings has been fully studied.

In fact, there exist other bosonic prototypes (see App.~B of this paper), but our discussion in the present series will be limited to the sets shown in Eq.~(1).

A central issue in all these theories is the presence of extra 
negative-norm states in the conformal field theories of the twisted sectors.  But the orbifold program constructs 
the twisted sectors of any orbifold from the untwisted sector and in 
string theory we do not expect that orbifoldization  would create negative-norm states where there are none 
in the untwisted sectors [21].  This observation implies that the twisted sectors of the new string theories
in fact possess new {\em extended world-sheet geometries}---including new twisted world-sheet permutation gravities [16], new twisted BRST systems [17], 
new extended physical-state conditions [17-20] and new gauges at the interacting level [19,20].  All these phenomena are associated with the existence of the 
so-called {\em orbifold Virasoro algebras} [1,2,9,16-20] (see Eqs.~(4.3) 
and (8.1b)), which appear universally in every sector of the orbifolds of permutation-type.  
It is possible , but so far demonstrated only in subexamples [19,20], that the new world-sheet geometries can eliminate all 
negative-norm states in the orbifold-string theories of permutation-type.

A closely-related issue is the {\em target space-time interpretation} of the new string theories, especially 
since detailed analysis of the physical states of the $\hc(\sigma) = 52$ twisted 
sectors has found an {\em equivalent, reduced formulation} [17] of all the $\hc(\sigma) = 52$  spectral 
problems at reduced central charge $c(\sigma) = 26$!
In the reduced formulation, one sees only (a unitary transformation of) the conventional number of negative-norm states subject in fact to 
the conventional physical-state condition.

Thus, for example, in spite of the half-integral moding of its $\hc = 52$ twisted open-string sector, the simplest orientation-orbifold string system with
\[
\bZ_2(w.s.) = (1,\tau_-),\ \ H_- = (1 \x (1)_{26};\tau_- \x (-1)_{26}) \eqno(1.2)
\]
has been shown to be equivalent [18], even at the interacting level [19], 
to the ordinary untwisted 26-dimensional open-closed string system.  In this sense, 
``orientation orbifolds include orientifolds'' [19] in a rising sequence of ever-more twisted open-closed string systems, with the familiar
untwisted critical prototype at the bottom 
of the hierarchy.  Similarly, the pure cyclic permutation-orbifold string 
systems $H_+ = \bZ_K$, $K =$ prime have been shown [20] to be equivalent to special 
modular-invariant collections of ordinary closed 26-dimensional strings, 
while more general $H'_{26}\subset H_{+}$ corresponds to 
generically new permutation-orbifold string theories.  In succeeding papers of this series, 
we will demonstrate that more general choices of $H'_{26} \subset 
H_{\pm}$  can in fact describe a large
variety of target spaces in the new theories, including in particular 
Lorentzian target space-times with sector-dependent dimensionality $D(\sigma) \le 26$!

Our task in this first paper of the new series is to begin the quantization of the full list of theories in Eq (1.1).  More precisely, 
we present here the BRST quantization [22,23] of the generalized permutation orbifold-string theories (1.1a) and their open-string counterparts (1.1b), all 
of whose sectors live at matter central charge $\hc(\s) = 26K$. (As noted 
above the twisted open-string sectors of the orientation-orbifold string 
theories are included in the case $K=2$, and in fact the BRST quantization of 
each closed-string in $U(1)^{26}/H'_{26}$ is quite ordinary.)
  
This paper in scope and in detail is therefore a generalization of the BRST quantization 
of $\hc = 52$ matter in Ref.~[17].  In particular, we follow Ref.~[17] in 
obtaining the general BRST quantization directly from the principle of local 
isomorphisms [3-15], leaving an equivalent derivation from the extended 
actions of the theories (see Ref.~16 and App.~B) for another time and place.

The mathematics needed for the generalization is minimal.  In orbifold 
theory, each twisted sector $\s$ corresponds to an 
equivalence class $\s$ of the divisor, and an appropriate language here is the familiar [7,9,16,18] {\em cycle-basis} for each
 sector $\s$ of the general permutation group $H(\perm)_K$:
\[
\hbj = 0,1,\dots,f_j(\s)-1,\ \ j = 0,1,\dots,N(\s) - 1,\ \ \sum_j f_j(\s) = K. \eqno(1.3)
\]
In this notation, $f_j(\s)$ and $N(\s)$ are respectively the length of cycle $j$ and the number of cycles in sector $\s$, while $\bhj$ 
indexes within each cycle.  For the convenience of the reader, the 
cycle-data for the elements of  $\bZ_K$ and $S_K$ are included in the text.  
At any stage of the development, the results of Ref.~[17] can be obtained from the cycle data
\[
K = 2,\ \ N(1) = 1,\ \ \hbj \to {\bar u} = 0,\ \ 1,\ \ f_0(1) = 2 \eqno(1.4)
\]
for the single non-trivial element $(\s = 1)$ of $H(\perm)_2 = \bZ_2$ or $\bZ_2(w.s.)$.

The results presented here include the twisted ghost and BRST systems of each cycle $j$ of each twisted sector $\s$ of
 all these orbifold-string systems, including in particular the {\em extended BRST algebra} of sector $\s$
\[
[\hQ_j(\s),\hQ_l(\s)]_+ = 0,\ \forall\ j,l \text{ in sector $\s$} \eqno(1.5)
\]
which was conjectured in Ref.~[17].  In a convention we follow throughout this paper, these are the BRST charges
of the open-string analogues (1.1c) only, and right-mover copies of each result must be added to describe
the generalized permutation orbifolds (1.1a).

The BRST systems then imply the {\em extended physical-state conditions} on the matter of cycle $j$ in sector $\s$
\[
( \hL_{\hj j}(( m + \frac {\hj}{f_j(\s)} ) \ge 0) - \ha_{f_j(\s)} \d_{m+ \frac {\hj}{f_j(\s)},0}) |\chi(\s)\>_j = 0 \eqno(1.6a)
\]
\[
\hc_j(\s) = 26f_j(\s),\ \ \ha_{f_j(\s)} = \frac {13f_j{}^2(\s)-1}{12f_j(\s)} \eqno(1.6b)
\] where $\{\hL_{\hj j}\}$ are the orbifold Virasoro generators (see Eq.~(4.3a)) with {\em cycle central charge} $\hc_j(\s)$
and the quantity $\ha_{f_j(\s)}$ is
called the {\em intercept of cycle $j$} in sector $\sigma$.  These results generalize those 
of Ref.~[17] $(\hat{c}=52$ and $\ha_2 = 17/8)$, and special cases of Eq.~(1.6) were also obtained at
the interacting level for $H_+ = H(\perm)_K = \bZ_K$, $K =$ prime (where 
the $K-1$ twisted sectors are single cycles with $f_{0}(\s) = K$).
The expected {\em sector central charges} of the matter
 \[ \hc(\s) = \sum_j \hc_j(\s) = 26K \eqno(1.7)\]
are obtained from Eqs.~(1.3) and (1.6b) by summing over the cycles of each sector.  
The extended physical-state conditions in Eq.~(1.6) are the quantum analogues of the classical extended Virasoro constraints $\{\hL_{\hj j} = 0\}$
obtained from the extended actions of these theories in Ref.~[16].

Because they descend universally from the twisted permutation gravities of Ref.~[16], the BRST systems presented here depend
only on the permutation groups $H(\perm)_K$ or $\bZ_2(w.s.)$.  The automorphism groups $H'_{26}$ in Eq.~(1.1) are however encoded in the explicit forms
of the orbifold Virasoro generators $\{\hL_{\hj j}\}$ of the matter, 
which will be constructed in the next paper of this series.  
There we will also discuss the equivalent, reduced formulation of the 
physical states of each cycle at reduced cycle central charge $c_j(\s) = 26$, and begin
 our study of the target space-time structure of the new string theories.

\section{BRST in the Orbifold Program}

We sketch here the standard steps of the orbifold program [3-15], which systematically constructs the operator-products
of the twisted sectors of any orbifold from the operator-products of the untwisted sector.

In this case we are interested in obtaining the BRST systems of the generalized permutation orbifolds and their open-string analogues
\[
\frac {U(1)^{26K}}{H_+},\ \ [ \frac {U(1)^{26K}}{H_+}]_{\open},\ \ H_+ \subset H(\perm)_K \x H'_{26} \eqno(2.1)
\]
so we begin with $K$ copies of the operator-product form of the untwisted BRST system given for $K = 2$ in Ref.~[17].  
This includes the composite operators
\[
J_I(z) =\ \lcol\bpsi_I(z)\psi_I(z)\rcol\,,\ \ I = 0,\dots,K-1 \eqno(2.2a)
\]
\[
\aligned
T_I^G(z) &= -\frac {1}{2} \lcol\bpsi_I(z) \overset{\leftrightarrow}{\p} \psi_I(z)\rcol + \frac {3}{2} \p J_I(z) \\
&\qquad =\ \lcol\bpsi_I(z)\p\psi_I(z)\rcol + 2\p\bpsi_I(z)\psi_I(z) \endaligned \eqno(2.2b)
\]
\[
J_I^B(z) = \bpsi_I(z)T_I(z) + \frac {1}{2} \lcol\bpsi_I(z)T_I^G(z)\rcol \eqno(2.2c)
\]
\[
T_I^t(z) = T_I(z) + T_I^G(z) \eqno(2.2d)
\]
as well as their operator products, e.g.,
\[
\psi_I(z)\bpsi_J(w) = \frac {\d_{IJ}}{(z-w)} + \lcol\psi_I(z)\bpsi_J(w)\rcol \eqno(2.3a)
\]
\[
\aligned
T_I^G(z)T_J^G(w) &= \d_{IJ} \{ \frac {-(26/2)}{(z-w)^4} + ( \frac {2}{(z-w)^2} + \frac {1}{z-w} \p_w) T_I^G(w)\} \\
&\qquad + \lcol T_I^G(z)T_J^G(w)\rcol 
\endaligned \eqno(2.3b)
\]
\[
\aligned
T_I(z)T_J(w) &= \d_{IJ} \{ \frac {(26/2)}{(z-w)^4} + ( \frac {2}{(z-w)^2} + \frac {1}{z-w} \p_w) T_I(w)\} \\
&\qquad + \lcol T_I(z)T_J(z)\rcol. 
\endaligned \eqno(2.3c)
\]
Here $\{\psi,\bpsi\}$ are K copies of the ghosts, taken as half-integer moded Bardakci-Halpern fermions [24], while $\{J\}$, $\{T^G\}$ 
and $\{J^{B}\}$ are the corresponding copies of the ghost current, the $c = -26$ ghost 
stress-tensor and the BRST current respectively.  The independent 
quantities $\{T\}$ are the copies of the $c = 26$ matter stress 
tensor while $\{T^t\}$ are the total stress-tensor copies, each with $c = 
0$. Finally, as seen in Eq.~(2.3), the symbol $\lcol\dots\rcol$ is 
operator-product normal ordering.

We remind the reader that here, and throughout the paper the steps are 
shown as above only for a single set of local operators, which is 
adequate to describe the open-string analogues in Eq.~(1.1c). A second 
set of right-mover operators is required to describe the closed strings 
of the generalized permutation orbifolds (1.1a). To obtain the twisted 
sectors of the orbifolds we will apply the principle of local 
automorphisms [3-15] (including monodromies), but for the open-string 
analogues this is only a simple shortcut known [15-18] to give the 
correctly-twisted open-string mode algebras \footnote{In fact of course 
twisted open strings generally have no monodromies, their fractional 
modeing being associated instead with the presence of different branes at 
each end of the string. See in particular Ref.~[15] where the general twisted
open WZW string and its branes are constructed from the left-movers of the
general WZW orbifold.}.

The automorphic response of all these fields $\{A\}$ to an arbitrary element of the general permutation group of order $K$ is
\[
A_I(z)' = \o(h_{\s})_I{}^J A_J(z),\ \ \o(h_{\s}) \in  H(\perm)_K. \eqno(2.4)
\]
The next step in the orbifold program is the construction for each operator $A$ the corresponding {\em eigenfield} $\cA$, 
whose automorphic response $\cA'$ to each $\o(h_{\s})$ is diagonal:
\[
\cA_{\hj j}(z,\s) \equiv \chi_{\hj j}(\s)U(\s)_{\hj j}{}^I A_I(z,\s) \eqno(2.5a)
\]
\[
\cA_{\hj j}(z,\s)' = e^{-2\pi i \frac {\hj}{f_j(\s)}} \cA_{\hj j}(z,\s) \eqno(2.5b)
\]
\[
\cA_{\hj \pm f_j(\s),j}(z,\s) = \cA_{\hj j}(z,\s) \eqno(2.5c)
\]
\[
\bhj = 0,\dots,f_j(\s)-1,\ \ j = 0,\dots,N(\s)-1 \eqno(2.5d)
\]
\[
\sum_j = N(\s),\ \ \sum_j f_j(\s) = K. \eqno(2.5e)
\]
Here $U(\s)$ is the unitary eigenmatrix of the eigenvalue problem of each element $\o(h_{\s}) \in H(\perm)_K$, which is 
known [7,9,16,18] in the {\em cycle-basis} of each element: $N(\s)$ is the total number of cycles $j$ in each $\o(h_{\s})$, while $\hj$ indexes 
within each cycle $j$ of length $f_j(\s) \ge 1$.  I have also included the standard periodicity convention (2.5c) within each 
cycle (which the eigenfields inherit from the eigenvalue problems) and $\bhj$ is the pullback of $\hj$ to its fundamental region.

In orbifold theory, each sector $\s$ corresponds to an equivalence class $\s$ of the automorphism group, so we need 
to choose one representative $\o(h_{\s})$ of each equivalence class of $H(\perm)_K$.  For example, each ordered partition of $K$
\[
\sum_j f_j(\s) = K,\ \ 1 \le f_{j+1}(\s) \le f_j(\s) \eqno(2.6)
\]
defines an equivalence class of the symmetric group $H(perm)_{K}=S_K$.  The cyclic group $H(\perm)_K = \bZ_K$ has $K$ sectors
\[
\o(h_{\s}) = e^{2\pi i \frac {\s}{K}},\ \ \s = 0,1,\dots,K-1,\ \ N(\s) = \frac {K}{f_j(\s)} \eqno(2.7)
\]
where $f_j(\s)$ is any divisor of $K$.  In the cyclic groups $f_j(\s)$ is also the order of each $\o(h_{\s})$.  
The results of Ref.~[16] can therefore be obtained from the results of this paper by choosing the non-trivial 
element $\s = 1$ of $H(\perm)_2 = \bZ_2$ with the single cycle-length $f_0(1) = 2$ (and $\bhj \to \bu = 0,1$).  More generally, 
the untwisted sector $\s = 0$ is described by the unit element
\[
\bhj = 0,\ \ f_j(0) = 1,\ \ N(0) = K \eqno(2.8)
\]
for all $H(\perm)_K$.

The quantities $\{\chi\}$ in the eigenfields (2.5a) are normalizations, which I choose here as
\[
\chi_{\hj j}(\s) = \begin{cases}
1 &\text{for ghosts} \\
f_j(\s) &\text{for BRST currents} \\
\sqrt{f_j(\s)} &\text{otherwise.}
\end{cases} \eqno(2.9)
\]
The last (square-root) convention here is the standard choice for 
permutation-orbifold matter [7,9,16-18] in the orbifold program, and the other conventions for the 
fermionic operators are consistent with the choice in Ref.~[17] for the non-trivial element of $H(\perm)_2 = \bZ_2$ with $f_j(\s) = 2$.  
It is then straightforward to compute the operator-products of the 
eigenfields in terms of the eigenfields themselves (see the remark after Eq.~(3.5)).

The final step in the orbifold program is an application of the {\em principle of local isomorphisms} [3-18], which maps 
the eigenfields $\{\cA\}$ of sector $\s$ to the twisted fields $\{\hA\}$ of sector $\s$:
\[
\cA_{\hat{j}j}(z,\s) \to \hA_{\hat{j}j}(z,\s). \eqno(2.10)
\]
There are two components to this principle.  First, the twisted fields 
are {\em locally isomorphic} to the eigenfields, that is, they have 
the same operator products.  Second, the {\em monodromies} of each twisted field $\hat{A}$
\[
\hA_{\hat{j}j}(ze^{2\pi i},\s) = e^{-2\pi i \frac {\hj}{f_j(\s)}} \hA_{\hat{j}j}(z,\s) \eqno(2.11)
\]
is the same as the diagonal automorphic response (2.5b) of the corresponding eigenfield $\cA$.  In this last step, 
and only here, do we change from the untwisted Hilbert space to the twisted Hilbert space of sector $\s$.  The final result 
for the operator-product form of the twisted BRST system is presented in the following section.

For perspective, we also mention an equivalent, alternate route (via the 
principle of local isomorphisms and a monodromy decomposition) to the 
same twisted fields  $\hat{A}$. This route and the one described above via 
eigenfields are both summarized in the so-called commuting diagrams 
[3,5,11] of the orbifold program.

\section{Operator-Product Form of the Twisted BRST's}

The twisted fields of cycle $j$ in sector $\s$ are as follows
\[
\hJ_{\hj j}(z,\s) = \sum_{\hl = 0}^{f_j(\s)-1} \lcol \hbpsi_{\hl j}(z,\s)\psi_{\hj-\hl,j}(z,\s)\rcol \eqno(3.1a)
\]
\[
\aligned
\hth_{\hj j}^G(z,\s) &= -\frac {1}{2} \sum_{\hl = 0}^{f_j(\s)-1} \lcol \hbpsi_{\hl j}(z,\s) \overset{\leftrightarrow}{\p} \hpsi_{\hj-\hl,j}(z,\s)\rcol + \frac {3}{2} \p \hJ_{\hj j}(z,\s) \\
&\qquad = \sum_{\hl=0}^{f_j(\s)-1} \lcol(\hbpsi_{\hl j}(z,\s) \p \hpsi_{\hj-\hl,j}(z,\s) + 2\p \hbpsi_{\hl j}(z,\s)\hpsi_{\hj-\hl,j}(z,\s))\rcol
\endaligned \eqno(3.1b)
\]
\[
\hJ_{\hj j}^B(z,\s) = \sum_{\hl = 0}^{f_j(\s)-1} \{\hbpsi_{\hl j}(z,\s)\hth_{\hj-\hl,j}(\s) + \frac {1}{2} \lcol\hbpsi_{\hl j}(z,\s)\hth_{\hj-\hl,j}^G(z,\s)\rcol\} \eqno(3.1c)
\]
\[
\hth_{\hj j}^t(z,\s) = \hth_{\hj j}(z,\s) + \hth_{\hj j}^G(z,\s) \eqno(3.1d)
\]
\[
\hA_{\hj j}(ze^{2\pi i},\s) = e^{-2\pi i \frac {\hj}{f_j(\s)}} \hA_{\hj j}(z,\s) \eqno(3.1e)
\]
\[
\hA_{\hj \pm f_j(\s),j}(z,\s) = \hA_{\hj j}(z,\s) \eqno(3.1f)
\]
\[
\bhj = 0,\dots,f_j(\s)-1,\ \ j = 0,\dots,N(\s)-1,\ \ \sum_j f_j(\s) = K \eqno(3.1g)
\]
where $\lcol\dots\rcol$ is still operator-product normal ordering and I have included the monodromies 
and periodicities of each twisted field in Eqs.~(3.1e) and (3.1f) 
respectively  The fields in this list include the twisted 
reparametrization ghosts $\{\hpsi,\hbpsi\}$, the twisted ghost currents $\{\hJ\}$, the extended ghost stress-tensors $\{\hth^G\}$ 
and the twisted BRST currents $\{\hJ^B\}$.  Finally $\{\hth\}$ and $\{\hth^t\}$ are respectively 
the extended matter stress-tensors and the extended total stress-tensors of cycle $j$ in sector $\s$.

The operator products of the twisted fields are also obtained in detail as follows.  We begin with the basic operator products 
of the twisted ghosts and their currents
\[
\hbpsi_{\hj j}(z,\s)\hpsi_{\hl l}(w,\s) = \frac {\d_{jl}\d_{\hj+\hl,0  \mod f_j(\s)}}{z-w} + \lcol\hbpsi_{\hj j}(z,\s)\hpsi_{\hl l}(w,\s)\rcol \eqno(3.2a)
\]
\[
\hbpsi_{\hj j}(z,\s)\hbpsi_{\hl l}(w,\s) =\ \lcol\hbpsi_{\hj j}(z,\s) \hbpsi_{\hl l}(w,\s)\rcol \eqno(3.2b)
\]
\[
\hpsi_{\hj j}(z,\s)\hpsi_{\hl l}(w,\s) =\ \lcol\hpsi_{\hj j}(z,\s)\hpsi_{\hl l}(w,\s)\rcol \eqno(3.2c)
\]
\[
\hJ_{\hj j}(z,\s)\hbpsi_{\hl l}(w,\s) = \frac {\d_{jl}\hbpsi_{\hj+\hl,j}(w)}{z-w} + \lcol\hJ_{\hj j}(z,\s)\hbpsi_{\hl l}(w,\s)\rcol \eqno(3.2d)
\]
\[
\hJ_{\hj j}(z,\s)\hpsi_{\hl l}(w,\s) = \frac {-\d_{jl}\hpsi_{\hj+\hl,j}(w)}{z-w} + \lcol\hJ_{\hj j}(z,\s)\hpsi_{\hl l}(w,\s)\rcol \eqno(3.2e)
\]
\[
\hJ_{\hj j}(z,\s)\hJ_{\hl l}(w,\s) = \frac {\d_{jl}\d_{\hj+\hl,0 \mod f_j(\s)}}{(z-w)^2} + \lcol\hJ_{\hj j}(z,\s)\hJ_{\hl l}(w,\s)\rcol \eqno(3.2f)
\]
and continue with the operator products involving the extended stress tensors:
\[
\aligned
\hth_{\hj j}(z,\s)\hth_{\hl l}(w,\s) &= \d_{jl}\{ \frac {( \frac {26}{2}) f_j(\s)\d_{\hj+\hl,0 \mod f_j(\s)}}{(z-w)^4} \\
&\qquad + ( \frac {2}{(z-w)^2} + \frac {1}{z-w} \p_w) \hth_{\hj+\hl,j}(w,\s)\} \\
&\qquad + \lcol\hth_{\hj j}(z,\s)\hth_{\hl l}(w,\s)\rcol
\endaligned \eqno(3.3a)
\]
\[
\aligned
\hth_{\hj j}^G(z,\s)\hth_{\hl l}^G(w,\s) &= \d_{jl} \{ \frac {-( \frac {26}{2}) f_j(\s)\d_{\hj+\hl,0\mod f_j(\s)}}{(z-w)^4} \\
&\qquad + ( \frac {2}{(z-w)^2} + \frac {1}{z-w} \p_w) \hth_{\hj+\hl,j}^G(w,\s)\} \\
&\qquad + \lcol\hth_{\hj j}^G(z,\s)\hth_{\hl l}^G(w,\s)\rcol
\endaligned \eqno(3.3b)
\]
\[
\aligned
\hth_{\hj j}^t(z,\s)\hth_{\hl l}^t(w,\s) &= \d_{jl} ( \frac {2}{(z-w)^2} + \frac {1}{z-w} \p_w) \hth_{\hj+\hl,j}^t(w,\s) \\
&\qquad + \lcol\hth_{\hj j}^t(z,\s)\hth_{\hl l}^t(w,\s)\rcol
\endaligned \eqno(3.3c)
\]
\[
\aligned
\hth_{\hj j}^G(z,\s)\hbpsi_{\hl l}(w,\s) &= \d_{jl} ( -\frac {1}{(z-w)^2} + \frac {1}{z-w} \p_w)\bpsi_{\hj+\hl,j}(w,\s) \\
&\qquad + \lcol\hth_{\hj j}^G(z,\s)\hbpsi_{\hl l}(w,\s)\rcol
\endaligned \eqno(3.3d)
\]
\[
\aligned
\hth_{\hj j}^G(z,\s)\hpsi_{\hl l}(w,\s) &= \d_{jl} ( \frac {2}{(z-w)^2} + \frac {1}{z-w} \p_w) \hpsi_{\hj+\hl,j}(w,\s) \\
&\qquad + \lcol\hth_{\hj j}^G(z,\s)\hpsi_{\hl l}(w,\s)\rcol
\endaligned \eqno(3.3e)
\]
\[
\aligned
\hth_{\hj j}^G(z,\s)\hJ_{\hl l}(w,\s) &= \d_{jl} ( \frac {1}{(z-w)^2} + \frac {1}{z-w} \p_w) \hJ_{\hj+\hl,j}(w) \\
&\qquad + \lcol\hth_{\hj j}^G(z,\s)\hJ_{\hl l}(w,\s)\rcol .
\endaligned \eqno(3.3f)
\]
Finally, we give the operator products which involve the twisted BRST currents:
\[
\aligned
\hJ_{\hj j}^B(z,\s)\hpsi_{\hl l}(w,\s) &= \d_{jl} ( \frac {\hJ_{\hj+\hl,j}(w,\s)}{(z-w)^2} + \frac {\hth_{\hj+\hl,j}^t(w,\s)}{z-w}) \\
&\qquad + \lcol\hJ_{\hj j}^B(z,\s)\hpsi_{\hl l}(w,\s)\rcol
\endaligned \eqno(3.4a)
\]
\[
\aligned
\hth_{\hj j}^t(z,\s)\hJ_{\hl l}^B(w,\s) &= \d_{jl} ( \frac {1}{(z-w)^2} + \frac {1}{z-w} \p_w) \hJ_{\hj+\hl,j}^B(w) \\
&\qquad + \lcol\hth_{\hj j}^t(z,\s)\hJ_{\hl l}^B(w,\s)\rcol
\endaligned \eqno(3.4b)
\]
\[
\aligned
\hJ_{\hj j}^B(z,\s)\hJ_{\hl l}^B(w,\s) &= \d_{jl}f_j(\s) \sum_{\hm = 0}^{f_j(\s)-1} \{ \frac {10}{(z-w)^3} \p_w \hbpsi_{\hm j}(w,\s)\hbpsi_{\hj+\hl-\hm,j}(w,\s) \\
&\qquad + \frac {5}{(z-w)^2} \p_w^2 \bpsi_{\hm j}(w,\s)\hbpsi_{\hj+\hl-\hm,j}(w,\s) \\
&\qquad + \frac {3}{2(z-w)} \p_w ( \p_w^2 \hbpsi_{\hm j}(w,\s)\hbpsi_{\hj+\hl-\hm,j}(w,\s))\} \\
&\qquad + \lcol\hJ_{\hj j}^B(z,\s)\hJ_{\hl l}^B(w,\s)\rcol
\endaligned \eqno(3.4c)
\]
\[
\bhj = 0,\dots,f_j(\s)-1,\ \ j = 0,\dots,N(\s)-1,\ \ \sum_j f_j(\s) = K. \eqno(3.4d)
\]
We should also mention that the extended matter stress tensors are independent of the ghost systems
\[
\hth_{\hj j}(z,\s)\hG_{\hl l}(w,\s) =\ \lcol\hth_{\hj j}(z,\s)\hG_{\hl l}(w,\s)\rcol \eqno(3.5)
\]
where $\hG$ can be $\hpsi,\hbpsi$ or any composite thereof.

Eqs.~(3.1) thru (3.4) are a complete description at the operator-product level of the twisted BRST systems of 
twisted sector $\s$.  The composite forms and operator products of the (intermediate) eigenfields $\{\cA\}$ can in fact be read 
from these equations by replacing $\hA \to \cA$ and the monodromies (3.1e) by the automorphic responses (2.5b).
Note also that the dynamics of twisted sector $\s$ is semisimple with respect to the cycles, as seen in the earlier 
conformal-field-theoretic study of the permutation orbifolds [9].

We close this section with some remarks about the operator-product algebra
of the extended stress tensors and their associated ordinary 
Virasoro subalgebras. From Eqs.~(3.3a-c), one reads the central charges of the extended stress-tensors of cycle $j$ in sector $\s$
\[
\hc_j(\s) = 26 f_j(\s) \text{ for } \{\hth\} \eqno(3.6a)
\]
\[
\hc_j^G(\s) = -26 f_j(\s) \text{ for } \{\hth^G\} \eqno(3.6b)
\]
\[
\hc_j^t(\s) = \hc_j(\s) + \hc_j(\s)^G = 0 \text{ for } \{\hth^t\}. \eqno(3.6c)
\]
In fact these ``cycle'' central charges are the central charges of the ordinary (Virasoro)
stress-tensors $\hth_{0j}$, $\hth_{0j}^G$ and $\hth_{0j}^t$ of cycle $j$, 
whose operator products have the schematic (Virasoro) form
\[
\aligned
\hth_{0j}(z,\s)\hth_{0l}(w,\s) &= \d_{jl} ( \frac {(\hat{c}_j(\s)/2)}{(z-w)^4} + \frac {2}{(z-w)^2} \p_w) \hth_{0j}(w) \\
&\qquad + \lcol\hth_{0j}(z,\s)\hth_{0l}(w,\s)\rcol
\endaligned \eqno(3.7)
\]
for each of the three types.  More familiar central charges are obtained for the {\em physical} stress tensors of sector $\s$
\[
\hth(z,\s) \equiv \sum_j \hth_{0j}(z,\s),\ \ \hth^G(z,\s) \equiv \sum_j \hth_{0j}(z,\s) \eqno(3.8a)
\]
\[
\hth^t(z,\s) \equiv \hth(z,\s) + \hth^G(z,\s) \eqno(3.8b)
\]
all three of which share the schematic (Virasoro) operator product:
\[
\aligned
\hth(z,\s)\hth(w,\s) &= \frac {\hc(\s)/2}{(z-w)^4} + ( \frac {2}{(z-w)^2} + \frac {1}{z-w} \p_w) \hth(w,\s) \\
&\qquad + \lcol\hth(z,\s)\hth(w,\s)\rcol
\endaligned \eqno(3.9a)
\]
\[
\hc(\s) = \sum_j \hc_j(\s) = 26K \eqno(3.9b)
\]
\[
\hc^G(\s) = \sum_j \hc_j^G(\s) = -26K \eqno(3.9c)
\]
\[
\hc^t(\s) =\,\, \hc(\s) + \hc^G(\s) \,= 0. \eqno(3.9d)
\]
The cycle-sums in (3.9b,c) were evaluated with the sum rule in Eq.~(3.1g), and we recognize
$26K$ in Eq.~(3.9b) as the matter central charge of each sector 
of any permutation orbifold on $K$ copies of $U(1)^{26}$.

\section{The Mode Algebras of Twisted Sector $\s$}

Consulting the monodromies (3.1e) and the conformal-weight terms $\{\D/(z-w)^2\}$ in the operator products with the extended 
stress tensors, we define the modes $\{\hA( m + \frac {\hj}{f_j(\s)} )\}$ of each twisted field as follows:
\[
\hbpsi_{\hj j}(z,\s) = \sum_{m \in \bZ} \hc_{\hj j}( m + \frac {\hj}{f_j(\s)} ) z^{-( m + \frac {\hj}{f_j(\s)}) + 1} \eqno(4.1a)
\]
\[
\hpsi_{\hj j}(z,\s) = \sum_{m \in \bZ} \hb_{\hj j}( m + \frac {\hj}{f_j(\s)}) z^{-( m + \frac {\hj}{f_j(\s)})-2} \eqno(4.1b)
\]
\[
\hJ_{\hj j}(z,\s) = \sum_{m \in \bZ} \hJ_{\hj j}( m + \frac {\hj}{f_n(\s)}) z^{-( m + \frac {\hj}{f_j(\s)})-1} \eqno(4.1c)
\]
\[
\hJ_{\hj j}^B(z,\s) = \sum_{m \in \bZ} \hJ_{\hj j}^B ( m + \frac {\hj}{f_j(\s)} ) z^{-( m + \frac {\hj}{f_j(\s)})-1} \eqno(4.1d)
\]
\[
\hth_{\hj j}^t(z,\s) = \sum_{m \in \bZ} \hL_{\hj j}^t ( m + \frac {\hj}{f_j(\s)}) z^{-( m + \frac {\hj}{f_j(\s)}) - 2} \eqno(4.1e)
\]
\[
\hL_{\hj j}^t ( m + \frac {\hj}{f_j(\s)}) = \hL_{\hj j} ( m + \frac {\hj}{f_j(\s)}) + \hL_{\hj j}^G( m + \frac {\hj}{f_j(\s)}) \eqno(4.1f)
\]
\[
\hA_{\hj \pm f_j(\s),j} ( m + \frac {\hj \pm f_j(\s)}{f_j(\s)} ) = \hA_{\hj j}( m \pm 1 + \frac {\hj}{f_j(\s)}) . \eqno(4.1g)
\]
The periodicity (4.1g) of the modes is a consequence of the periodicity (3.1f) of the twisted fields.

Then by standard contour methods [5] we find the mode algebra of  
cycle $j$ in each twisted sector $\s$, beginning with 
the twisted ghosts and their currents
\[
[ \hc_{\hj j}( m + \frac {\hj}{f_j(\s)} ), \hb_{\hl l} ( n + \frac {\hl}{f_l(\s)})]_+ = \d_{jl} \d_{m + n + \frac {\hj+\hl}{f_j(\s)},0} \eqno(4.2a)
\]
\[
[ \hc_{\hj j}( m + \frac {\hj}{f_j(\s)}), \hc_{\hl l} ( n + \frac {\hl}{f_l(\s)})]_+ = [\hb_{\hj j}(m + \frac {\hj}{f_j(\s)}), \hb_{\hl l}(n + \frac {\hl}{f_l(\s)})]_+ = 0 \eqno(4.2b)
\]
\[
[ \hJ_{\hj j}( m + \frac {\hj}{f_j(\s)}), \hc_{\hl l} ( n + \frac {\hl}{f_l(\s)})] = \d_{jl} \hc_{\hj+\hl,j} ( m + n + \frac {\hj+\hl}{f_j(\s)}) \eqno(4.2c)
\]
\[
[ \hJ_{\hj j} ( m + \frac {\hj}{f_j(\s)}), \hb_{\hl l} ( n + \frac {\hl}{f_l(\s)})] = -\d_{jl} \hb_{\hj+\hl,j} ( m + n + \frac {\hj+\hl}{f_j(\s)}) \eqno(4.2d)
\]
\[
[ \hJ_{\hj j} ( m + \frac {\hj}{f_j(\s)}), \hJ_{\hl l} ( m + \frac {\hl}{f_l(\s)})] = \d_{jl} f_j(\s) ( m + \frac {\hj}{f_j(\s)}) \d_{m+n+\frac {\hj+\hl}{f_j(\s)},0} \eqno(4.2e)
\]
\[
\bhj = 0,\dots,f_j(\s) - 1,\ \ j = 0,\dots,N(\s) - 1,\ \ \sum_j f_j(\s) = K. \eqno(4.2f)
\]
One consequence of our normalization choice (2.9) is the simple ghost anticommutator (4.2a).

We turn next to the mode algebras of the extended stress tensors, which 
include the following three extended Virasoro algebras:  
\[
\aligned
&[ \hL_{\hj j} ( m + \frac {\hj}{f_j(\s)}), \hL_{\hl l} ( n + \frac {\hl}{f_l(\s)})] \\
&\qquad = \d_{jl} \{ ( m - n + \frac {\hj-\hl}{f_j(\s)}) \hL_{\hj+\hl,j} ( m + n + \frac {\hj+\hl}{f_j(\s)}) \\
&\qquad + \frac {26f_j(\s)}{12} ( m + \frac {\hj}{f_j(\s)}) ( ( m + \frac {\hj}{f_j(\s)})^2 - 1) \d_{m+n+\frac {\hj+\hl}{f_j(\s)},0} \}
\endaligned \eqno(4.3a)
\]
\[
\aligned
&[ \hL_{\hj j}^G ( m + \frac {\hj}{f_j(\s)}), \hL_{\hl l}^G ( n + \frac {\hl}{f_l(\s)})] \\
&\qquad = \d_{jl} \{ ( m - n + \frac {\hj-\hl}{f_j(\s)} ) \hL_{\hj+\hl,j}^G ( m + n + \frac {\hj+\hl}{f_j(\s)})  \\
&\qquad - \frac {26f_j(\s)}{12} ( m + \frac {\hj}{f_j(\s)} ) ( ( m + \frac {\hj}{f_j(\s)})^2 - 1) \d_{m+n+\frac {\hj+\hl}{f_j(\s)},0} \} 
\endaligned \eqno(4.3b)
\]
\[
\aligned
&[ \hL_{\hj j}^t ( m + \frac {\hj}{f_j(\s)}), \hL_{\hl l}^t ( n + \frac {\hl}{f_l(\s)}) ] \\
&\qquad = \d_{jl} ( m - n + \frac {\hj+\hl}{f_j(\s)}) \hL_{\hj+\hl,j}^t ( m + n + \frac {\hj+\hl}{f_j(\s)})
\endaligned \eqno(4.3c)
\]
\[
[ \hL_{\hj j} ( m + \frac {\hj}{f_j(\s)}), \hL_{\hl l}^G ( n + \frac {\hl}{f_l(\s)})] = 0. \eqno(4.3d)
\]
These algebras are also called general {\em orbifold Virasoro algebras} 
[1,2,9,16-20], and in particular the  algebra of the matter generators 
$\{\hat{L}_{\hat{j}j}\}$ at fixed $j$ is an orbifold 
Virasoro algebra of order $f_{j}(\sigma)$.  Each of the three orbifold Virasoro 
algebras contains a so-called {\em integral Virasoro subalgebra} for each 
cycle $j$ in sector $\sigma$
\[
\hth_{0j}(z,\s) = \sum_{m \in \bZ} \hL_{0j}(m)z^{-m-2} \eqno(4.4a)
\]
\[
[\hL_{0j}(m),\hL_{0l}(n)] = \d_{jl}\{ (m-n) \hL_{0j}(m+n) + \frac {\hc_j(\s)}{12} m(m^2-1) \d_{m+n,0}\} \eqno(4.4b)
\]
\[
\hc_j(\s) = \begin{cases}
26f_j(\s) &\text{for $\{\hL_{0j}\}$} \\
-26f_j(\s) &\text{for $\{\hL_{0j}^G\}$} \\
0 &\text{for $\{\hL_{0j}^t\}$}
\end{cases} \eqno(4.4c)
\]
which exhibit the three cycle central charges discussed in the previous 
section. Because the cycle dynamics is semisimple, there are also three 
physical Virasoro subalgebras for each sector $\s$
\[
\hat{\th}(z,\s) = \sum_j \hL_{\s}(m)z^{-m-2} \eqno(4.5a)
\]
\[
\hL_{\s}(m) \equiv \sum_j \hL_{0j}(m) \eqno(4.5b)
\]
\[
[\hL_{\s}(m),\hL_{\s}(n)] = (m-n) \hL_{\s}(m+n) + \frac {\hc(\s)}{12} m(m^2-1)\d_{m+n,0} \eqno(4.5c)
\]
\[
\hc(\s) = \sum_j \hc_j(\s) = \begin{cases}
26K &\text{for $\{\hL_{\s}\}$} \\
-26K &\text{for $\{\hL_{\s}^G\}$} \\
0 &\text{for $\{\hL_{\s}^t\}$.}
\end{cases} \eqno(4.5d)
\]
which exhibit the expected sector central charges. 

Of course, the extended Virasoro generators of the matter are independent of the ghost systems
\[
[\hL_{\hj j} ( m + \frac {\hj}{f_j(\s)}), \hG_{\hl l} ( n + \frac {\hl}{f_l(\s)})] = 0 \eqno(4.6)
\]
where $\hG$ can be $\hc,\hb$ or any composite thereof. This statement includes Eq.~(4.3d) as a special case.

Other algebras in the twisted ghost systems include the following:
\[
\aligned
&[ \hL_{\hj j}^G ( m + \frac {\hj}{f_j(\s)}), \hc_{\hl l} ( n + \frac {\hl}{f_l(\s)})] \\
&\qquad = -\d_{jl} ( 2 ( m + \frac {\hj}{f_j(\s)}) + n + \frac {\hl}{f_l(\s)}) \hc_{\hj+\hl,j} ( m + n + \frac {\hj+\hl}{f_j(\s)})
\endaligned \eqno(4.7a)
\]
\[
\aligned
&[ \hL_{\hj j}^G ( m + \frac {\hj}{f_j(\s)}), \hb_{\hl l} ( n + \frac {\hl}{f_l(\s)})] \\
&\qquad = \d_{jl} ( m - n + \frac {\hj+\hl}{f_j(\s)}) \hb_{\hj+\hl,j} ( m + n + \frac {\hj+\hl}{f_j(\s)})
\endaligned \eqno(4.7b)
\]
\[
\aligned
&[ \hL_{\hj j}^G ( m + \frac {\hj}{f_j(\s)}), \hJ_{\hl l} ( n + \frac {\hl}{f_l(\s)})] \\
&\qquad = \d_{jl} ( m - n + \frac {\hj+\hl}{f_j(\s)}) \hJ_{\hj+\hl,j} ( m + n + \frac {\hj+\hl}{f_j(\s)}).
\endaligned \eqno(4.7c)
\]
Finally, we have the algebra which involves the modes of the twisted BRST currents:
\[
[ \hL_{\hj j}^t ( m + \frac {\hj}{f_j(\s)}), \hJ_{\hl l}^B ( n + \frac {\hl}{f_l(\s)})] = -\d_{jl} ( n + \frac {\hl}{f_j(\s)}) \hJ_{\hj+\hl,j}^B ( m + n + \frac {\hj+\hl}{f_j(\s)}) \eqno(4.8a)
\]
\[
\aligned
&[ \hJ_{\hj j}^B ( m + \frac {\hj}{f_j(\s)}), \hb_{\hl l} ( n + \frac {\hl}{f_l(\s)})]_+ \\
&\qquad = \d_{jl}\{ ( m + \frac {\hj}{f_j(\s)}) \hJ_{\hj+\hl,j} ( m + n + \frac {\hj+\hl}{f_j(\s)}) + \hL_{\hj+\hl,j}^t ( m + n + \frac {\hj+\hl}{f_j(\s)})\}
\endaligned \eqno(4.8b)
\]
\[
\aligned
&[ \hJ_{\hj j}^B ( m + \frac {\hj}{f_j(\s)}), \hJ_{\hl l}^B ( n + \frac {\hl}{f_l(\s)})]_+ \\
&\qquad = \d_{jl} f_j(\s) \{ 5 ( m + \frac {\hj}{f_j(\s)}) ( n + \frac {\hl}{f_j(\s)}) -  \frac {3}{2} ( m + n + \frac {\hj+\hl}{f_j(\s)}) ( m + n + \frac {\hj+\hl}{f_j(\s)} - 1)\} \\
&\qquad \x \sum_{\hm = 0}^{f_j(\s)-1} \sum_{p \in \bZ} ( p + \frac {\hm}{f_j(\s)}) \hc_{\hm j} ( p + \frac {\hm}{f_j(\s)}]) \hc_{\hj+\hl-\hm,j} ( m + n - p + \frac {\hj+\hl-\hm}{f_j(\s)})
\endaligned \eqno(4.8c)
\]
\[
\bhj = 0,\dots,f_j(\s)-1,\ \ j = 0,\dots,N(\s)-1,\ \ \sum_j f_j(\s) = K. \eqno(4.8d)
\]
Appendix~A includes a technical remark used to simplify the right side of the anticommutator (4.8c) of the BRST current modes.  
Also, evaluation of this anticommutator at $f_j(\s) = 2$ shows a (non-propagating) typo in Eq (4.4i) of Ref.~[17], where the 
last factor before the sums should read $( m + n + \frac {u+v}{2} - 1)$ instead of $( m + n + \frac {u+v}{2} + 1)$.

We conclude with the operator-product normal-ordered forms of the composite mode operators themselves
\[
\hJ_{\hj j} ( m + \frac {\hj}{f_j(\s)}) = \sum_{\hl = 0}^{f_j(\s)-1} \sum_{p \in \bZ} \lcol\hc_{\hl j} ( p + \frac {\hl}{f_j(\s)}) \hb_{\hj-\hl,j} ( m - p + \frac {\hj-\hl}{f_j(\s)})\rcol \eqno(4.9a)
\]
\[
\aligned
&\hL_{\hj j}^G ( m + \frac {\hj}{f_j(\s)}) \\
&\qquad = -\sum_{\hl=0}^{f_j(\s)-1} \sum_{p \in \bZ} ( m + p + \frac {\hj+\hl}{f_j(\s)}) \lcol\hc_{\hl j} ( p + \frac {\hl}{f_j(\s)} ) \hb_{\hj+\hl,j} ( m - p + \frac {\hj-\hl}{f_j(\s)})\rcol
\endaligned \eqno(4.9b)
\]
\[
\aligned
&\hJ_{\hj j}^B ( m + \frac {\hj}{f_j(\s)}) \\
&\qquad = \sum_{\hl = 0}^{f_j(\s)-1} \sum_{p \in \bZ} \{ \hc_{\hl j} ( p + \frac {\hl}{f_j(\s)}) \hL_{\hj-\hl,j} ( m - p + \frac {\hj-\hl}{f_j(\s)})\, + \\
&\qquad \quad \ \ \ \ \   +  \frac {1}{2} \lcol\hc_{\hl j} ( p + \frac {\hl}{f_j(\s)}) \hL_{\hj-\hl,j}^G ( m - p + \frac {\hj-\hl}{f_j(\s)})\rcol\}
\endaligned \eqno(4.9c)
\]
but these forms are not as useful as the mode-ordered forms of these operators given in Sec.~7.

\section{One BRST Charge per Cycle per Sector}

In the discussion above, we have constructed the modes $\{\hat{J}^{B}_{\hat{j}j}(m+\hat{j}/f_{j}(\sigma))\}$ of one twisted 
BRST current per cycle per 
sector for the general bosonic orbifold-string system of 
permutation-type. We may then define exactly one BRST charge per cycle per sector
\[
\hQ_j(\s) \equiv \hJ_{0j}^B(0),\ \ j = 0,1,\dots,N(\s) - 1 \eqno(5.1)
\]
as the unique zero mode $(m=\hat{j}=0)$ of each twisted BRST current.

As a consequence, the results above imply the following algebra of the BRST charges with the other operators in sector $\s$
\[
[ \hQ_j(\s),\hL_{\hl l}^t ( n + \frac {\hl}{f_l(\s)} ) ] = 0,\ \forall\ j,l,\s \eqno(5.2a)
\]
\[
[ \hQ_j(\s), \hb_{\hl l} ( n + \frac {\hl}{f_l(\s)})]_+ = \d_{jl} \hL_{\hl j}^t ( n + \frac {\hl}{f_j(\s)}) \eqno(5.2b)
\]
\[
\aligned
&[ \hQ_j(\s), \hJ_{\hl l}^B ( n + \frac {\hl}{f_l(\s)})]_+ \\
&\qquad = -\d_{jl} \frac {3}{2} f_j(\s) ( n + \frac {\hl}{f_j(\s)}) ( n + \frac {\hl}{f_j(\s)} - 1 ) \\
&\qquad \quad \ \ \ \ \x \sum_{\hj = 0}^{f_j(\s)-1} \sum_{p \in \bZ} ( p + \frac {\hj}{f_j(\s)}) \hc_{\hj j} ( p + \frac {\hj}{f_j(\s)} ) \hc_{\hl-\hj,j} ( n - p + \frac {\hl-\hj}{f_j(\s)})
\endaligned \eqno(5.2c)
\]
as well as the previously conjectured algebra [17] of the BRST charges themselves:
\[
[\hQ_j(\s),\hQ_l(\s)]_+ = 0,\ \forall\ j,l,\s \eqno(5.3a)
\]
\[
\hQ_j^2(\s) = 0. \eqno(5.3b)
\]
The extended BRST algebra in Eq.~(5.3a) follows directly from Eq.~(5.2c), 
and comprises one of the central results of this paper.

In addition to the implied nilpotency (5.3b) of each BRST charge, I note the following two consistency checks among the 
algebras (5.2), (5.3),(4.3c),(4.6) and (4.7b).  Both checks involve using Eq.~(5.2b) at $j = l$ as a definition of the total
orbifold Virasoro generators $\{\hL^{t}_{\hl j}\}$.  First, adding to this the BRST algebra (5.3a), we find that the algebra (5.2a) is implied.  Second, 
an independent derivation of the total orbifold Virasoro algebra (4.3c) is obtained as 
follows: Start on the left side of Eq.~(4.3c) and use (5.2b) for one of the total Virasoro generators in the commutator.
Then  sequential application of the algebras (5.2b), (4.6), (4.7b) and a final application of  
Eq.~(4.2b) gives the right side of Eq.~(4.3c). These are of course 
generalizations of the consistency checks of the ordinary, 
untwisted BRST system (see e.g. Ref.~[23]) and the twisted BRST system of 
$\hat{c}=52$ matter [17].

\section{The Physical States of Cycle $j$ in Sector $\s$}

For each sector $\s$ of each orbifold of permutation-type, we use the BRST charges (5.1) 
to define the {\em physical states} $\{|\chi(\s)\>_j\}$ of cycle $j$ as follows:
\[
\hQ_j(\s) |\chi(\s)\>_j = 0 \eqno(6.1a)
\]
\[
\hb_{\hj j}( ( m + \frac {\hj}{f_j(\s)}) \ge 0) |\chi(\s)\>_j = 0 \eqno(6.1b)
\]
\[
\hc_{\hj j}( ( m + \frac {\hj}{f_j(\s)} ) > 0 ) |\chi(\s)\>_j = 0 \eqno(6.1c)
\]
\[
\bhj = 0,\dots,f_j(\s) - 1,\ \ j = 0,\dots,N(\s)-1,\ \ \sum_j f_j(\s) = K. \eqno(6.1d)
\]
Using then Eqs.~(5.2b) and (6.1a,b), we find also that the physical 
states are also annihilated by the non-negative modes of the total 
orbifold Virasoro generators
\[
\hL_{\hj j}^t ( ( m + \frac {\hj}{f_j(\s)} ) \ge 0) |\chi(\s)\>_j = 0 \eqno(6.2)
\]
which is consistent because these generators have zero central extension.  The role of 
the $\hc$-condition in Eq.~(6.1c) will be noted in the following section.

Summing on the cycles, we find that the physical states 
are also annihilated by the total BRST charge of sector $\s$
\[
\hQ(\s) \equiv \sum_j \hQ_j(\s),\ \ \hQ^2(\s) = 0 \eqno(6.3a)
\]
\[
\hQ(\s)|\chi(\s)\> = 0,\ \ |\chi(\s)\> \equiv \otimes_j |\chi(\s)\>_j \eqno(6.3b)
\]
but this condition is less useful than the BRST conditions for each cycle in Eq.~(6.1a).

\section{Mode-Ordered Form of the Twisted Ghost Systems}

To further analyze the physical states, we need more explicit forms of the operators in the twisted ghost systems.  
For this purpose, I define the following {\em mode normal-ordered product}
\[
\aligned
&\lcol\hA_{\hj j} ( m + \frac {\hj}{f_j(\s)}) \hB_{\hl l} ( n + \frac {\hl}{f_l(\s)})\rcol_M \\
&\qquad \equiv -\th(( m + \frac {\hj}{f_j(\s)}) > 0) \hB_{\hl l} ( n + \frac {\hl}{f_l(\s)}) \hA_{\hj j}( m + \frac {\hj}{f_j(\s)}) \\
&\qquad + \frac {1}{2} \d_{m + \frac {\hj}{f_j(\s)},0} [ \hA_{0j}(0),\hB_{\hl l} ( m + \frac {\hl}{f_l(\s)}) ] \\
&\qquad + \th( ( m + \frac {\hj}{f_j(\s)}) < 0) \hA_{\hj j} ( m + \frac {\hj}{f_j(\s)}) \hB_{\hl l} ( n + \frac {\hl}{f_l(\s)})
\endaligned \eqno(7.1)
\]
where $\hA$ and $\hB$ can be either $\hc$ or $\hb$.

Then the mode expansions in Eq.~(4.1) straightforwardly give the following relation between the 
operator-product normal-ordered quadratics and the mode normal-ordered quadratics
\[
\lcol\hbpsi_{\hj j}(z,\s)\hpsi_{\hl l}(w,\s)\rcol\ =\ \lcol\hbpsi_{\hj j}(z,\s)\hpsi_{\hl l}(w,\s)\rcol_M + \d_{jl} \d_{\hj+\hl,0 \mod f_j(\s)} ( \hD_{\bhj j}(z,w) - \frac {1}{z-w} ) \eqno(7.2a)
\]
\[
\aligned
&\hD_{\bhj j}(z,w) \equiv ( \frac {z}{w})^{2-\frac {\bhj}{f_j(\s)}} \frac {1}{z-w} - \frac {z}{2w^2} \d_{\bhj,0} \\
&\qquad = \frac {1}{z-w} + \ha_{\bhj j}^{(0)}(w) + (z-w)\ha_{\bhj j}^{(1)}(w) + 0(z-w)^2
\endaligned \eqno(7.2b)
\]
\[
\ha_{\bhj j}^{(0)}(w) \equiv \frac {1}{w} ( 2 - \frac {\bhj}{f_j(\s)} - \frac {1}{2} \d_{\bhj,0}) \eqno(7.2c)
\]
\[
\ha_{\bhj j}^{(1)}(w) \equiv \frac {1}{2w^2} \{ ( 2 - \frac {\bhj}{f_j(\s)}) ( 1 - \frac {\bhj}{f_j(\s)}) - \d_{\bhj,0}\} \eqno(7.2d)
\]
where $\bhj = 0,1,\dots,f_j(\s) - 1$ is again the pullback of $\hj$ to the fundamental region.  Comparing to the exact expression
\[
\hD_{0j} (z,w) = \frac {z(z+w)}{2w^2(z-w)} \eqno(7.3)
\]
we see that the terms given explicitly in the Laurent expansion above are exact for $\bhj = 0$.

Eq.~(7.2) then implies the local relations
\[
\lcol\hbpsi_{\hj j}(z,\s)\hpsi_{\hl l}(z,\s)\rcol\ =\ \lcol\hbpsi_{\hj j}(z,\s) \hpsi_{\hl l}(z,\s)\rcol_M + \d_{jl} \d_{\hj+\hl,0 \mod f_j(\s)} \ha_{\bhj j}^{(0)}(z) \eqno(7.4a)
\]
\[
\lcol\hbpsi_{\hj j}(z,\s) \p_z \hpsi_{\hl l}(z,\s)\rcol\ =\ \lcol\hbpsi_{\hj j}(z,\s) \p_z \hpsi_{\hl l}(z,\s)\rcol_M + \d_{jl}\d_{\hj+\hl,0 \mod f_j(\s)}( \p_z\ha_{\bhj j}^{(0)}(z) - \ha_{\bhj j}^{(1)}(z)) \eqno(7.4b)
\]
\[
\lcol\p_z \hbpsi_{\hj j}(z,\s) \hpsi_{\hl l}(z,\s)\rcol\ =\ \lcol\p_z \hbpsi_{\hj j}(z,\s) \hpsi_{\hl l}(z,\s)\rcol_M + \d_{jl}\d_{\hj+\hl,0 \mod f_j(\s)} \ha_{\bhj j}^{(1)}(z) \eqno(7.4c)
\]
which are sufficient to obtain the mode-ordered forms of the relevant operators of the ghost system.

We begin with the mode-ordered form of the twisted ghost currents $\{\hJ^G\}$:
\[
\aligned
\hJ_{\hj j}^G(z,\s) &\equiv \hJ_{\hj j}(z,\s) - \frac {3f_j(\s)}{2z} \d_{\hj,0 \mod f_j(\s)} \\
&\qquad = \sum_{\hl = 0}^{f_j(\s)-1} \lcol\hbpsi_{\hl j}(z,\s) \hpsi_{\hj-\hl,j}(z,\s)\rcol_M \\
&\qquad = \sum_{m \in \bZ} \hJ_{\hj j}^G ( m + \frac {\hj}{f_j(\s)}) z^{-( m + \frac {\hj}{f_j(\s)})-1}
\endaligned \eqno(7.5a)
\]
\[
\aligned
\hJ_{\hj j}^G( m + \frac {\hj}{f_j(\s)}) &= \hJ_{\hj j}( m + \frac {\hj}{f_j(\s)}) - \frac {3f_j(\s)}{2} \d_{m + \frac {\hj}{f_j(\s)},0} \\
&= \sum_{\hl = 0}^{f_j(\s)-1} \sum_{p \in \bZ} \lcol\hc_{\hl j}( p + \frac {\hl}{f_j(\s)}) \hb_{\hj-\hl,j}( m - p + \frac {\hj-\hl}{f_j(\s)})\rcol_M .
\endaligned \eqno(7.5b)
\]
In what follows we use only these mode-ordered forms $\{\hJ^G\}$ as the properly-ordered ghost currents.  Since they 
differ from the original ghost currents $\{\hJ\}$ only by a $c$-number 
shift, we may in fact read $\hJ \to \hJ^G$ 
in all the mode algebras of Section 4.  Using 
the definition (7.1) of the mode-ordering, we can for example write out the {\em twisted ghost charge of cycle $j$} in full detail:
\[
\aligned
\hJ_{0j}^{G}(m = 0) &= \frac {1}{2} [\hc_{0j}(0),\hb_{0j}(0)] + \sum_{p=1}^{\i} (\hc_{0j}(-p)\hb_{0j}(p) - \hb_{0j}(-p)\hc_{0j}(p)) \\
&\qquad - \sum_{\hl = 1}^{f_j(\s)-1} \hb_{-\hl,j} ( - \frac {\hl}{f_j(\s)}) \hc_{\hl j} ( \frac {\hl}{f_j(\s)}) \\
&\qquad + \sum_{\hl = 1}^{f_j(\s)-1} \sum_{p=1}^{\i} \{ \hc_{\hl j} ( -p + \frac {\hl}{f_j(\s)}) \hb_{-\hl,j} ( p - \frac {\hl}{f_j(\s)}) \\
&\qquad\qquad \ \ \ \ \ \ \   -  \hb_{-\hl,j} ( -p - \frac {\hl}{f_j(\s)}) \hc_{\hl j}( p + \frac {\hl}{f_j(\s)})\}.
\endaligned \eqno(7.6)
\]
Note that the terms involving the zero modes $\hc_{0j}(0)$ and $\hb_{0j}(0)$ of the twisted ghosts are isomorphic to 
the standard ghost-charge operator of ordinary (untwisted) BRST.

With Eqs.~(3.1b) and (7.2), we also obtain the mode-ordered form of the extended ghost stress-tensors and their orbifold Virasoro generators:
\[
\aligned
&\hth_{\hj j}^G(z,\s) = -\frac {\ha_{f_j(\s)}}{z^2} \d_{\hj,0\mod f_j(\s)} \\
&\qquad\ \ \ \ \ \ \ + \sum_{\hl = 0}^{f_j(\s)-1} \lcol(\hbpsi_{\hl j}(z,\s) \p_z \hpsi_{\hj-\hl,j}(z,\s) + 2\p_z \hbpsi_{\hl j}(z,\s) \hpsi_{\hj-\hl,j}(z,\s))\rcol_M
\endaligned \eqno(7.7a)
\]
\[
\aligned
&\hL_{\hj j}^G (m + \frac {\hj}{f_j(\s)}) = -\ha_{f_j(\s)} \d_{m+\frac {\hj}{f_j(\s)},0}\\
&\qquad\ \ \ \ \ \ \  - \sum_{\hl = 0}^{f_j(\s)-1} \sum_{p \in \bZ} ( m + p + \frac {\hj+\hl}{f_j(\s)}) \lcol \hc_{\hl j} ( p + \frac {\hl}{f_j(\s)}) \hb_{\hj-\hl,j} ( m - p + \frac {\hj-\hl}{f_j(\s)})\rcol_M
\endaligned \eqno(7.7b)
\]
\[
\ha_{f_j(\s)} \equiv \frac {13f_j^2(\s) - 1}{12f_j(\s)}. \eqno(7.7c)
\]
In what follows, I shall refer to the quantity $\ha_{f_j(\s)}$ in Eq.~(7.7c) as the {\em intercept of cycle $j$} in each sector $\s$.  
Mode-ordered expressions can also be obtained for the twisted BRST currents $\{\hJ_{\hj j}^{G}\}$ and the BRST charges $\{\hQ_j\}$, but these 
will not be needed in the present development.

As an application of the mode-ordered forms (7.5b) and (7.7b), we may use {\em both} the $\hb$ and the $\hc$ 
conditions (6.1b,c) on the physical states of cycle $j$ to compute directly that
\[
( \hJ_{\hj j}^G ( ( m + \frac {\hj}{f_j(\s)} ) \ge 0) + \frac {1}{2} \d_{m + \frac {\hj}{f_j(\s)},0}) |\chi(\s)\>_j = 0 \eqno(7.8a)
\]
\[
(\hL_{\hj j}^G(( m + \frac {\hj}{f_j(\s)} ) \ge 0) + \ha_{f_j(\s)} \d_{m + \frac {\hj}{f_j(\s)},0}) |\chi(\s)\>_j = 0. \eqno(7.8b)
\]
We emphasize that these results require attention to the definition (7.1) of mode-ordering.  In particular, the explicit 
form (7.6) of the twisted ghost charge is needed to see that the physical 
states of cycle $j$ have ghost number $( -\frac {1}{2})$, uniformly, as 
in ordinary untwisted BRST (see e.g. Ref.~[25])).

For completeness, we finally give the action on the physical states of 
the ghost current modes and ghost Virasoro generators of sector $\s$
\[
\hJ_{\s}^G(m) \equiv \sum_j \hJ_{0j}^G(m),\ \ \hL_{\s}^G(m) \equiv \sum_j \hL_{0j}^G(m) \eqno(7.9a)
\]
\[
(\hJ_{\s}^G(m \ge 0) + \frac {N(\s)}{2} \d_{m,0}) |\chi(\s)\> = 0 \eqno(7.9b)
\]
\[
(\hL_{\s}^G(m \ge 0) + \frac {1}{12} ( 13K - \sum_j \frac {1}{f_j(\s)}) \d_{m,0}) |\chi(\s)\> = 0 \eqno(7.9c)
\]
where $\{L_{\s}^G(m)\}$ satisfies the Virasoro algebra (4.5c) at $\hc^G(\s) = -26K$, $N(\s)$ is the 
number of cycles in sector $\s$ and the  product states $\{|\chi(\s)\>\}$ are defined in Eq (6.3b).

\section{The Extended Physical-State Conditions}

The final step in this paper is the transition from the action of the ghost operators to that of the {\em matter operators} on
the physical states.

This transition is now effected in a single step, using Eq (7.8b) and the fact (6.2) that the physical states are annihilated 
by the non-negative modes of the total orbifold Virasoro generators $\{\hL^t\}$ with $\hc_j^t(\s) = 0$.  The result is the 
following {\em extended physical-state conditions of cycle $j$ in sector $\s$}:
\[
( \hL_{\hj j}( ( m + \frac {\hj}{f_j(\s)}) \ge 0) - \ha_{f_j(\s)} \d_{m + \frac {\hj}{f_j(\s)},0} ) |\chi(\s)\>_j = 0 \eqno(8.1a)
\]
\[
\aligned
&[ \hL_{\hj j} ( m + \frac {\hj}{f_j(\s)} ), \hL_{\hl l} ( l + \frac {\hl}{f_l(\s)})] \\
&\qquad = \d_{jl} \{ ( m - n + \frac {\hj-\hl}{f_j(\s)}) \hL_{\hj+\hl,j} ( m + n + \frac {\hj+\hl}{f_j(\s)})  \\
&\qquad\ \ \ \ \ \  +  \frac {\hc_j(\s)}{12} ( m + \frac {\hj}{f_j(\s)}) ( ( m + \frac {\hj}{f_j(\s)})^2 - 1) \d_{m+n+\frac {\hj+\hl}{f_j(\s)},0}\}
\endaligned \eqno(8.1b)
\]
\[
\hc_j(\s) = 26f_j(\s),\ \ \ha_{f_j(\s)} = \frac {13f_j^2(\s) - 1}{12f_j(\s)} \eqno(8.1c)
\]
\[
\bhj = 0,1,\dots,f_j(\s) - 1,\ \ j = 0,1,\dots,N(\s)-1,\ \ \sum_j f_j(\s) = K. \eqno(8.1d)
\]
These conditions on the twisted matter of the new string theories are a central result of the paper. We remind 
that the extended Virasoro algebra (8.1b) of each cycle $j$ is 
called an orbifold Virasoro algebra of order $f_{j}(\sigma)$ at cycle central 
charge $\hc_j(\s) = 26f_j(\s)$, and the {\em intercept} $\ha_{f_j(\s)}$ of cycle $j$ in
sector $\s$ descends directly from Eq.~(7.8b) of the ghost system. 

The extended physical state conditions (8.1a) are the operator analogues of the classical extended Virasoro constraints
\[
\hth_{\hj j}(z,\s) = \hL_{\hj j}( m + \frac {\hj}{f_j(\s)}) = 0,\,\, \forall\ \hj j \text{ in sector } \s \eqno(8.3)
\]
associated to the general permutation gravities of the extended actions 
given for these theories in Ref.~[16]. Further remarks on the 
action formulations are included in App.~B.

In this sense the cycle central charges $\{\hc_j(\s)\}$ and the cycle-intercepts $\{\ha_{f_j(\s)}\}$ are the fundamental numbers 
obtained in our BRST quantization, and both quantities are universal in that they depend only on the length $f_j(\s)$ of cycle $j$ in 
sector $\s$.  The following table of numerical values
\[
\begin{tabular}{c|r|c}
$f_j(\s)$ & $\hc_j(\s)$ & $\ha_{f_j(\s)}$ \\ \hline
1 & 26~ & 1 \\
2 & 52~ & ~17/8~ \\
3 & 78~ & ~29/9~ \\
4 & 104~ & ~69/16 \\
5 & 130~ & ~27/5~ \\
6 & 156~ & 155/24
\end{tabular}
\]
underscores that both quantities increase monotonically with cycle-length.

Using the cycle basis of each element $\o(h_{\s}) \in H(\perm)_K$, the twisted matter 
of each sector $\s$ of each orbifold-string of permutation-type is 
described by cycle-collections of our result in Eq.~(8.1).  
I illustrate this here with some simple examples.

The first case in the table $f_j(\s) = 1$ includes ordinary untwisted 
critical string theory.  Copies of ordinary string theory are encountered here for example in 
the untwisted sectors $\s = 0$ (the trivial element of $H(\perm)_K$ and $H'_{26}$) of all the orbifold-string theories of permutation-type:
\[
(\hL_{0j}(m \ge 0) - \d_{m,0}) |\chi(0)\>_j = 0 \eqno(8.4a)
\]
\[
[\hL_{0j}(m),\hL_{0l}(n)] = \d_{jl} ( (m-n)L_{0j}(m+n) + \frac {\hc_j(0)}{12} m(m^2-1) \d_{m+n,0}) \eqno(8.4b)
\]
\[
\hc_j(0) = 26,\ \ N(0) = K,\ \ \bhj = 0,\ \ j = 0,1,\dots,K-1. \eqno(8.4c)
\]
With a set of right-mover copies and the relabeling $j \to I$, this is 
recognized as the physical-state conditions of the original $K$ copies $U(1)^{26K}$ of the ordinary 
untwisted closed string $U(1)^{26}$, i.e., the starting point of the orbifold program in this case.  There are no cycles of unit length 
in the non-trivial twisted sectors $\s = 1,\dots,K-1$ of $H(\perm)_K = \bZ_K$, but cycles 
of unit length occur 
frequently  in the elements of $H(\perm)_K =S_K$ -- where they also describe 
unit-cycle strings with $\hc_{j}(\sigma)=26$ and unit intercept.  We emphasize 
however that these 
unit-cycle strings are not ordinary strings for non-trivial elements of the 
space-time automorphism group $H'_{26}$. Indeed, for example, 
each sector $\sigma$ of the ordinary space-time orbifolds 
$U(1)^{26}/H'_{26}$ (the closed-string sectors of each 
orientation-orbifold string system) is also described by the ordinary 
physical-state condition (8.4) at $K=1$.

The simplest example of the second case $f_j(\s) = 2$ is the single 
twisted sector $\s = 1$ of $H(\perm)_2 = \bZ_2$ or $\bZ_{2}(w.s.)$, where we find
\[
( \hL_{\hj0}( m + \frac {\hj}{2} ) - \frac {17}{8} \d_{m+\frac {\hj}{2},0}) |\chi(1)\>_0 = 0 \eqno(8.5a)
\]
\[
\aligned
&[ \hL_{\hj 0}( m + \frac {\hj}{2}), \hL_{\hl 0} ( n + \frac {\hl}{2})] \\
&\qquad = ( m - n + \frac {\hj-\hl}{2}) \hL_{\hj+\hl,0}( m + n + \frac {\hj+\hl}{2}) \\
&\qquad \ \ \ \ \ + \frac {52}{12} ( m + \frac {\hj}{2}) ( ( m + \frac {\hj}{2})^2 - 1) \d_{m+n+\frac {\hj+\hl}{2},0}
\endaligned \eqno(8.5b)
\]
\[
\hat{c}(1) = \hat{c}_{0}(1) = 52, \,\,\,N(1) = 1,\ \ j = 0,\ \ \bhj = 0,1. \eqno(8.5c)
\]
With the relabeling $\bhj \to \bu = 0,1$, this collection is recognized 
as the two extended physical-state conditions of twisted $\hc = 52$ matter 
in Refs.~[16-20]. These  systems appear for example in the twisted open-string sectors of the orientation-orbifold 
string theories [15-19] or, adding right-mover copies, in the 
generalized $\bZ_2$-permutation orbifolds [15-18,20].

The result (8.5) is included in the extended physical-state conditions for the twisted 
sectors $\s = 1,\dots,K-1$ of $\bZ_K$, $K =$ prime, where $f_j(\s) = K$ for each sector:
\[
( \hL_{\hj 0 }( ( m + \frac {\hj}{K}) \ge 0) - \ha_K\d_{m + \frac {\hj}{K},0}) |\chi(\s)\>_0 = 0 \eqno(8.6a)
\]
\[
\aligned
&[ \hL_{\hj 0} ( m + \frac {\hj}{K}), \hL_{\hl 0} ( n + \frac {\hl}{K}) ] \\
&\qquad = (m - n + \frac {\hj-\hl}{K}) \hL_{\hj+\hl,0} ( m + n + \frac {\hj+\hl}{K}) \\
&\qquad \ \ \ \ \ + \frac {\hc_0(\s)}{12} ( m + \frac {\hj}{K}) ( ( m + \frac {\hj}{K})^2 - 1) \d_{m+n+\frac {\hj+\hl}{K},0}
\endaligned \eqno(8.6b)
\]
\[
\hc_0(\s) = 26K,\ \ \ha_K = \frac {13K^2-1}{12K},\ \ K = \text{prime} \eqno(8.6c)
\]
\[
N(\s) = 1,\ \ j = 0,\ \ \bhj = 0,1,\dots,K-1,\ \ \s = 1,\dots,K-1. \eqno(8.6d)
\]
The extended physical-state conditions for prime cyclic groups were conjectured in Ref.~[18] and 
verified at the interacting level in Ref.~[20].

The cycle-bases for the $K-1$ twisted sectors of the general cyclic group $H(\perm)_K = \bZ_K$ 
\[
\bhj = 0,1,\dots,f_j(\s)-1,\ \ j = 0,1,\dots, \frac {K}{f_j(\s)}-1,\ \ \sum_j f_j(\s) = K \eqno(8.7a)
\]
\[
\s = 1,\dots,K-1 \eqno(8.7b)
\]
were described in Eq.~(2.7). Then for example we find the following extended physical-state conditions for the three 
twisted sectors $\s = 1,2,3$ of $H(\perm)_4 = \bZ_4$:  There are two single-cycle sectors with length 4
\[
( \hL_{\hj 0}( ( m + \frac {\hj}{4}) \ge 0) - \frac {69}{16} \d_{m + \frac {\hj}{4},0}) |\chi(\s)\>_0 = 0 \eqno(8.8a)
\]
\[
\aligned
&[ \hL_{\hj 0} ( m + \frac {\hj}{4}), \hL_{\hl 0} ( n + \frac {\hl}{4})] \\
&\qquad = ( m - n + \frac {\hj-\hl}{4}) \hL_{\hj+\hl,0} ( m + n + \frac {\hj+\hl}{4}) \\
&\qquad\ \ \ \ \  + \frac {104}{12} ( m + \frac {\hj}{4} ) ( ( m + \frac {\hj}{4})^2 - 1) \d_{m+n+\frac {\hj+\hl}{4},0}
\endaligned \eqno(8.8b)
\]
\[
j = 0,\ \ \bhj = 0,1,2,3,\ \ \s = 1,3 \eqno(8.8c)
\]
and another sector with two cycles of length 2
\[
( \hL_{\hj j} ( ( m + \frac {\hj}{2} ) \ge 0 ) - \frac {17}{8} \d_{m + \frac {\hj}{2},0} ) |\chi(2)\>_j = 0 \eqno(8.9a)
\]
\[
\aligned
&[\hL_{\hj j}(m+\frac {\hj}{2}),\hL_{\hl l}(n + \frac {\hl}{2})] \\ 
&\qquad \ \ \ \ \ = \d_{jl} \{ ( m - n + \frac {\hj-\hl}{2} ) \hL_{\hj-\hl,j} ( m + n + \frac {\hj+\hl}{2})  \\
&\qquad \ \ \ \ \ \ \ \ \ \ +  \frac {52}{12} ( m + \frac {\hj}{2}) ( ( m +\frac {\hj}{2})^2 - 1) \d_{m+n+\frac {\hj+\hl}{2},0}\}
\endaligned \eqno(8.9b)
\]
\[
j = 0,1,\ \ \bhj = 0,1,\ \ \s = 2. \eqno(8.9c)
\]
For $H(\perm)_8 = \bZ_8$, one has a single-cycle sector with length 8, five sectors with two cycles of length 4 each, 
and one sector with four cycles of length 2.  Further cyclic examples are 
left to the reader, and we remind that the cycle-bases for the permutation 
groups $H(\perm)_{K}=S_{K}$ are given in Eq.~(2.6).

Finally, we mention the physical-state condition of any sector $\s$ (see Eq.~(4.5))
\[
\hL_{\s}(m) = \sum_j \hL_{0j}(m) \eqno(8.10a)
\]
\[
( \hL_{\s}(m \ge 0) - \frac {\d_{m,0}}{12} ( 13K - \sum_j \frac {1}{f_j(\s)})) |\chi(\s)\> = 0 \eqno(8.10b)
\]
\[
[\hL_{\s}(m),\hL_{\s}(n)] = (m-n) \hL_{\s}(m+n) + \frac {26K}{12} m(m^2-1) \d_{m+n,0} \eqno(8.10c)
\]
where $\{\hL_{\s}\}$ are the physical Virasoro generators of sector $\s$ 
for any $H(\perm)_K$ and the product-states   $\{|\chi(\s)\>\}$ are 
defined in Eq.~(6.3b). With the sum rule in Eq.~(8.1d), this condition follows simply from our central result (8.1) by summing over the cycles of each sector.

\section{Conclusions}

We have completed a BRST quantization of the bosonic prototypes of the generalized permutation-orbifold string theories and their open-string analogues
\[
\frac {U(1)^{26K}}{H_+},\ \ [ \frac {U(1)^{26K}}{H_+} ]_{\open},\ \ H_+ 
\subset H(\perm)_K \x H'_{26} \eqno(9.1)
\]
where $H'_{26}$ is any automorphism group of the critical closed string $U(1)^{26}$ and $H(\perm)_K$ is any
permutation group on the $K$ copies. The matter of each sector $\sigma$ 
of these orbifolds lives at sector central charge $\hat{c}(\sigma)=26K$. 
We remind that the orientation-orbifold string systems
\[
\frac {U(1)^{26}}{H_-} = \frac {U(1)_L^{26} \x U(1)_R^{26}}{H_-},\ \ H_- \subset \bZ_2(w.s.) \x H'_{26} \eqno(9.2)
\]
are also described in our quantization: The twisted open-string sectors of 
these theories are included in the open-string analogues at $K=2$ and 
$\hat{c}(\sigma)=52$, while the twisted closed-string sectors are the ordinary 
space-time orbifolds $U(1)^{26}/H_{26}'$ at $\hat{c}(\sigma)=26$. These last 
cases correspond to the trivial element of $\bZ_2(w.s.)$ -- and hence 
ordinary  physical-state conditions for each sector. 
The extended action 
formulations [16] of the bosonic prototypes are reviewed in App.~B, 
which also points out a generalization of the orientation-orbifold string systems.

Using the cycle-bases  of general permutation groups, our central result 
in this paper is the construction of
one twisted BRST system per cycle per sector in each of the 
orbifold-string systems (9.1) and (9.2). 
In particular, we have found the  {\em extended BRST algebra} (5.3a), 
conjectured in Ref.~(17), which gives one nilpotent 
BRST charge $\hQ_j(\s)$ for each cycle $j$ in each sector $\s$ of these orbifolds.  The twisted BRST systems then imply 
the {\em extended physical state conditions} (8.1) for the matter at {\em cycle central charge} $\hc_j(\s) = 26f_j(\s)$, 
where $f_j(\s)$ is the length of cycle $j$. The sector central charges of 
the matter are 
recovered as the cycle sums $\hat{c}(\sigma)=\sum_{j}\hat{c}_{j}(\sigma)=26K$. The extended
physical-state conditions also exhibit another set of fundamental numbers, 
the so-called {\em cycle-intercepts} $\ha_{f_j(\s)}$ in Eq.~(8.1c). We 
remind that a right-mover copy of the results given here are necessary to 
describe the twisted closed-string sectors of these orbifolds.

Our  results here therefore generalize the construction in Ref.~(17) of one twisted BRST system
for the non-trivial element of $H(perm)_{2}=\bZ_{2}$ or $\bZ_{2} (w.s.)$, with  
single-cycle length $f_{0}(1)= 2$ and  matter central charge $\hat{c}_{0}(1)=52$.
We emphasize that all aspects of our new twisted BRST systems, including 
the extended physical-state conditions, are {\em separable} in the cycles $j$ of
 each sector.  This is in accord with the extended actions [16] of these theories (see App.~B), and with earlier 
 work on the WZW permutation orbifolds [9] with trivial $H'_{26}$.  

In this paper, the generators $\{ \hL_{\hj j} ( m + \frac {\hj}{f_j(\s)})\}$ of the orbifold Virasoro algebras 
of the matter have been treated abstractly, so our first task in the next paper of this series 
will be the explicit construction of these generators in terms of the 
twisted matter fields. It is in this step that each element of the 
automorphism group $H'_{26}$ is encoded. With this information, 
we will also find the equivalent, {\em reduced} formulation of the physical-state problem of each  cycle $j$ at 
{\em reduced cycle-central charge} $c_j(\s) = 26$, and begin our study of the {\em target space-times} of 
the orbifold-string theories of permutation-type.

\clearpage
\section*{Appendix A.  Fermi Statistics for $\hc$'s}

For any numerical function $\cO$, define the quadratic inner product
\[
(\hc,\cO\hc)_{\hj j}(m) \equiv \sum_{\hl = 0}^{f_j(\s)-1} \sum_{p \in \bZ} \hc_{\hl j} ( p + \frac {\hl}{f_j(\s)}) \hc_{\hj-\hl,j} ( m - p + \frac {\hj-\hl}{f_j(\s)}) \cO \eqno(A.1)
\]
on the twisted ghost modes $\hc$.  Then the vanishing anticommutator (4.2b) of any two $\hc$'s gives 
the following list of equivalences among various pairs of $\cO$'s
\[
1 \approx 0 \eqno(A.2a)
\]
\[
p^2 \approx mp \eqno(A.2b)
\]
\[
( p + \frac {\hl}{f_j(\s)})^2 \approx ( m + \frac {\hj}{f_j(\s)}) ( p + \frac {\hl}{f_j(\s)}) \eqno(A.2c)
\]
in this inner product, and these equivalences were used to simplify the right-side of Eq (4.8c).

\clearpage
\section*{Appendix B.  Extended Actions and \\
${}$\hskip 1 in \ \ \ \ \ \ \ \   Generalized Orientation Orbifolds}

Ref.~[16] gave the extended actions for the $\hc = 52$ twisted open-string sectors of the orientation-orbifold string systems
\[
\frac {U(1)_L^{26} \x U(1)_R^{26}}{H_-},\ \ H_- \subset \bZ_2(w.s.)\x H'_{26} \eqno(B.1a)
\]
\[
\hS_{\s} = \frac {1}{4\pi} \int dt \int_0^{\pi} d\xi \sum_{u,v,w = 0}^1 \hh_{(u+v+w)}^{mn} \hH^{(u)} \p_m \hX^{n(r)\mu v} \cG_{n(r)\mu;n(s)\nu}(\s) \p_m \hX^{n(s)\nu w}. \eqno(B.1b)
\]
Here $\hat{h}_{(u)}^{mn}$ is the inverse of $\hh_{mn}^{(u)}$, which is the twisted metric of $\bZ_2$-permutation gravity associated to the world-sheet 
orientation-reversing element of $\bZ_2(w.s.)$.  The explicit forms of the extended diffeomorphisms, 
the densities $\hH^{(u)}$ and the twisted metric $\cG(\sigma)$ of $H'_{26}$ are also given there, as well as the branes of these strings.  
The generators $\{ \hL_u ( m + \frac {u}{2} ) \}$ of the orbifold 
Virasoro algebras  
are singly-twisted and the matter currents $\{ \hJ ( m + \frac 
{n(r)}{\rho(\s)} + \frac {u}{2})\}$ are doubly-twisted, where 
$\{n(r)/\rho(\sigma)\}$ are the spectral fractions of the elements of $H'_{26}$. 
The twisted $\hc = 26$ closed-string sectors of the orientation orbifold form the ordinary space-time 
orbifold $U(1)^{26}/H'_{26}$, with ordinary Polyakov gravity, ordinary Virasoro algebras and singly-twisted currents $\{ \hJ ( m + \frac {n(r)}{\rho(\s)})\}$.

Ref.~[16] also gave the extended actions of each twisted $\hc = 26K$ closed-string sector of the generalized permutation orbifolds
\[
\frac {U(1)^{26K}}{H_+},\ \ H_+ \subset H(\perm)_K \x H'_{26} \eqno(B.2a)
\]
\[
\aligned
\hS_{\s} &= \frac {1}{8\pi} \int dt \int_0^{2\pi} d\xi \sum_j f_j(\s) \sum_{\hj,\hh,\hl = 0}^{f_j(\s)-1}\x \\
&\qquad \ \ \ \ \ \x \hh_{(\hj+\hk+\hl)j}^{mn} \hH^{(\hj)j} \p_m \hX^{n(r)\mu\hk j} \cG_{n(r)\mu;n(s)\nu}(\s) \p_n \hX^{n(s)\nu\hl j}
\endaligned \eqno(B.2b)
\]
\[
\sum_j f_j(\s) = K \eqno(B.2c)
\]
where $h_{mn}^{(\hj)j}$ is the extended metric of the permutation gravity associated to each equivalence class of $H(\perm)_K$.  The monodromies, 
extended diffeomorphisms and the explicit form of the densities 
$\hH^{(\hj)j}$ in these sectors are also given in Ref.~[16].  
Again, the generators $\{ \hL_{\hj j} ( m + \frac {\hj}{f_j(\s)})\}$ of the orbifold Virasoro algebra are singly-twisted 
and the matter currents $\{ \hJ ( m + \frac {n(r)}{\rho(\s)} + \frac {\hj}{f_j(\s)})\}$ are doubly-twisted.  In these cases 
of course one obtains a left- and right-mover copy of each algebra.

On the basis of these results, one expects that the extended actions of the $\hc = 26K$ open-string analogues 
of the generalized permutation orbifolds will have the same form as that shown 
in  Eq.~(B.2), but with the substitution
\[
[ \frac {U(1)^{26K}}{H_+} ]_{\open}: \int_0^{2\pi} d\xi \to \int_0^{\pi} d\xi. \eqno(B.3)
\]
By construction these extended actions have the same bulk-invariances and modeing as 
those of the generalized permutation orbifolds, and these actions reduce for $K = f_j(\s) = 2$ to the actions (B.1) of the open-string sectors
of the orientation orbifolds. See also the construction of general 
twisted open strings in Ref.~[15].

Note that the extended actions (B.1-3) are {\em separable} in the 
cycle-label $j$, 
consistent with the results of the text and earlier work in Refs.~[9,16-18].  In principle, 
the twisted BRST systems of this paper can also be derived from these actions by the Faddeev-Popov method.

In fact, there exist other sets of bosonic prototypes of the new string theories, for example the {\em generalized orientation-orbifold} string theories
\[
\frac {(U(1)_L^{26} \x U(1)_R^{26})^K}{H_{-+}},\ \ H_{-+} \subset \bZ_2(w.s.) \x H(\perm)_K \x H'_{26} \eqno(B.4)
\]
which reduce to the orientation orbifolds (B.1a) when $K = 1$.  Detailed 
construction of these theories for higher $K$ is beyond the scope of this paper, 
so I confine the discussion here to some preliminary remarks on their structure.

I begin with some simple  properties of these theories as CFT's.  Like orientation orbifolds, the presence of the world-sheet 
orientation group $\bZ_2(w.s.)$ in the divisor tells us 
that the generalized orientation orbifolds have an equal number of twisted open- and closed-string sectors, 
now at sector central charges $\hc = 52K$ and $26K$ respectively.  In 
fact, the closed-string sectors (corresponding to the trivial element of 
$\bZ_2(w.s.)$) of these theories  clearly form (two copies of) the generalized permutation 
orbifold (B.2a), while the sectors corresponding to the trivial 
element of $H(\perm)_K$ comprise (K copies of) the ordinary orientation orbifold (B.1a).

It is also clear that the matter currents of the open-string sectors of 
these theories are {\em triply-twisted}, with three spectral fractions 
corresponding to the elements of $H_{26}', H(perm)_{K}$ and $\bZ_{2}(w.s.)$:
\[
\hJ( m + \frac {n(r)}{\rho(\s)} + \frac {\hj}{f_j(\s)} + \frac {u}{2}) \eqno(B.5a)
\]
\[
\bn(r) \in [0,1,\dots,\rho(\s)-1],\ \ \bu = 0,1 \eqno(B.5b)
\]
\[
\bhj = 0,1,\dots,f_j(\s)-1,\ \ j = 0,1,\dots,N(\s)-1,\ \ \sum_j f_j(\s) = K. \eqno(B.5c)
\]
Triply-twisted matter is a degree of complexity not yet studied in the orbifold program.  

By the same token, the open-string sectors should exhibit new {\em doubly-twisted 
orbifold-Virasoro generators} of the form $\{ \hL_{\hj ju}( m + \frac {\hj}{f_j(\s)} + \frac {u}{2})\}$, as well as doubly-twisted 
BRST systems $(\hc_{\hj ju},\hb_{\hj ju},\hJ_{\hj ju}^B,\dots)$ and 
open-string extended actions of the schematic form:
\[
\aligned
\hS_{\s} &= \frac {1}{8\pi} \int dt \int_0^{\pi} d\xi \sum_j f_j(\s) 
\sum_{\hj,\hk,\hl = 0}^{f_j(\s)-1} \sum_{u,v,w = 0}^1 \x \\
&\qquad \ \ \ \ \ \x \hh_{(\hj+\hk+\hl)j(u+v+w)}^{mn} \hH^{(\hj)j(u)} \p_m \hX^{n(r)\mu\hk jv} \cG_{n(r)\mu;n(s)\nu}(\s) \p_n \hX^{n(s)\nu\hl jw}.
\endaligned \eqno(B.6)
\]
Here the quantities $\hat{h}_{mn}^{(\hat{j})j(u)}$ and 
$\hat{H}^{(\hat{j})j(u)}$ are the extended metric and density of the 
world-sheet permutation gravity associated with each equivalence class of 
the product permutation group $\bZ_{2}(w.s.)\x H(perm)_{K}$. I have not 
worked out these last structures in detail, and beyond this I will not speculate in this paper.

\section*{Acknowledgements}

For helpful discussion and encouragement, I thank L.~Alvarez-Gaum\'e, C.~Bachas, J.~de~Boer, S.~Frolov, O.~Ganor, E.~Kiritsis, 
A.~Neveu, H.~Nicolai, N.~Obers, B.~Pioline, M.~Porrati, E.~Rabinovici, V.~Schomerus, C.~Schweigert, M.~Staudacher, R.~Stora,
C.~Thorn, E.~Verlinde and J.-B.~Zuber.

\end{document}